\pdfoutput=1
\documentclass[aps,pre, twocolumn, groupedaddress]{revtex4-1} 
\usepackage{lipsum}
\usepackage{mathtools}
\usepackage{graphicx}

\usepackage{amsmath}    
\usepackage{amssymb}
\usepackage{bm}
\usepackage{hyperref}
\usepackage{latexsym}
\usepackage{verbatim}
\usepackage{color}
\usepackage[caption=false]{subfig}

\def\beq{\begin{equation}}
\def\eeq{\end{equation}}

\def\beq{\begin{equation}}                          
\def\eeq{\end{equation}}                          
\def\bea{\begin{eqnarray}}                          
\def\eea{\end{eqnarray}}

\DeclareRobustCommand{\uvec}[1]{{%
  \ifcsname uvec#1\endcsname
     \csname uvec#1\endcsname
   \else
    \bm{\hat{\mathbf{#1}}}%
   \fi
}}

\preprint{}

\begin{document}

\title{Directional cues affect the collective behaviour of Self propelled  particles in one dimension}

\author{Pawan Kumar Mishra}
\email{pawankumarmishra.rs.phy19@itbhu.ac.in}
\affiliation{Indian Institute of Technology (BHU) Varanasi, India 221005}
\author{Abhra Puitandy}
\email{abhrapuitandy.rs.phy22@itbhu.ac.in}
\affiliation{Indian Institute of Technology (BHU) Varanasi, India 221005}
\author{Shradha Mishra}
\email[]{smishra.phy@itbhu.ac.in}
\affiliation{Indian Institute of Technology (BHU) Varanasi, India 221005}
\date{\today}

\begin{abstract}
{This study explores the effect of quenched disorder on the characteristic of self-propelled particles in one-dimension. Here,particles interact with disorder  which serve as directional cues. The study investigates how the density of the disorder influence the emergence of ordering and clustering  in the collection  of the self propelled particles. We introduce the microscopic model  as well as corresponding coarse-grained equations of motion for the local density and the orientation of particle.  Disorder affects the macroscopic ordering in the system, the size of the ordered clusters decays algebraically with disorder. Further, the disorder also affects the clustering of particles; in the presence of disorder, a big macroscopic cluster breaks into small clusters, leads to the localization of particles around it and results in high density around the disorder.}
\end{abstract}

\maketitle

\section{Introduction}
Collective behaviour of active agents is an interesting area of research in recent years\cite{active_1,active_2,active_3,active_4,active_5,active_6,active_7,dikshit2023ordering}.  The example of such systems can be observed in a wide range of length scales\cite{volpe2022active}. Starting from   very small scale: like cytoskeleton filaments inside our cell \cite{needleman2017active}, bacterial colony \cite{fodor2018statistical} to larger scales like; bird flock \cite{feder2007statistical}, school of fish \cite{partridge1982structure}, human crowd \cite{bottinelli2017jammed}, animal herd \cite{lundmark2015one} etc.  Many artificially designed systems in lab are also good examples to understand the characteristics of the active systems\cite{narayan2007long}. These systems show interesting nonequilibrium phase transition, steady state  and dynamical properties: like, long-ranged ordering in two-dimensions, large density fluctuations, enhanced dynamics, faster information propagation which is in general not present in the corresponding equilibrium systems \cite{mishra2022active,nishiguchi2023deciphering, abp_1,abp_2,abp_3,abp_4}.\\

Most of the previous  studies on dynamic and steady state  properties of active matter systems, focusses on systems in clean or homogeneous medium \cite{tonertupre1998, Toner1995, Toner2005, pattanayak2018collection}. But disorder or inhomogeneity are inevitable in nature \cite{mishrainhomogeneous, morin2017distortion, chepizhko2013optimal, yllanes2017many,reichhardt2017disorder}. Some of the recent studies have shown the effect of disorder on the ordered state of polar self-propelled particles in two and three dimensions \cite{rakeshpre, morin2017distortion,chepizhko2013optimal,yllanes2017many,quint2015topologically,sandor2017pre, singh2021bond, singh2021ordering, pattanayak2020speed, kumar2020active,kumar2021effect,kumar2022active, das2020nonquenched,das2018ordering}. It is observed that disorder can break the robust long ranged ordered state in two-dimensions and changes it to quasi-long ranged order. Whereas in three-dimensions, the ordering remains unaffected by the presence of disorder. \\
The presence of long-ranged ordered state in clean systems is also observed not only in two and higher dimensions but in one-dimension as well. The natural way of observing such long-ranged ordering in one dimension can be a moving ant trail \cite{attygalle1985ant,goss1990trail}, or traffic flow in a single lane \cite{herman1959single,wang2007synchronized}. The long-range ordering of the one-dimensional wires is observed in quasi-1D system\cite{Enforced_LRO_1d}.   This motivated  us to find the role of disorder for the collection of active agents in one-dimension. We strictly keep the nature of disorder such that it only influence the direction of incoming particles, hence they are like field disorder or {\em directional cues}. Any other type of physical disorder will block the path of particles and immediately break the clustering and ordering among the particles. \\
We model our system as a collection of active agents moving on a one-dimensional ring (due to periodic boundary condition) with short ranged alignment interaction with the neighbouring particles as well as the random field disorder present in the system. The position and direction of disorder remains quenched in time. When an incoming trail of particles encounter a disorder, it tries to get reoriented in the direction of disorder. If the size of the trail is large, and the direction of disorder is opposite to its incoming direction, then a part of it can get reflected and some part can be transmitted. This is scenario when disorder density is small. As we increase disorder density, then a moving particle cluster encounter with disorder more frequently and it leads to the breaking of a big cluster into many small clusters. Many times these small clusters are simply stuck between two oppositely oriented disorder. Hence disorder also becomes responsible for local trapping of moving cluster. For all disorder densities the two-point orientation correlation function shows a mixture of algebraic and exponential decay. The  cluster size algebraically decreases  with the density of disorder. System shows good dynamical scaling for all disorder densities. Results obtained from microscopic simulation and coarse-grained model are in good match with each other.\\


\section{Model}
{\em Microscopic Model:-}
We study a collection of self-propelled particles with discrete symmetry in one-dimension in the presence of disorder. 
Disorders, which are placed random in space and quenched in time act like directional cues. 
We have considered $N$($N_d$) particles(disorders) placed randomly along a line of length $L$. Hence the number density of particles(disorders) $\rho$$(\rho_d)$ = $\frac{N(N_d)}{L}$, where  $N_d << N$. The disorder also have the same symmetry as particles, but their direction is quenched in time. The position and orientation of $i^{th}$ particle  at time $t$ are given by $x_{i}(t)$ and dimensionless velocity $u_{i}(t)$ with randomly chosen initial value $\pm1$. The orientation of the disorder is also randomly fixed to $\pm1$ in space and quenched in time. The position and orientation of the particle is updated by
\begin{equation}
x_i(t+1)=x_i(t)+v_0 u_i(t)
\end{equation}
\begin{equation}
u_i(t+1)=G({\langle u(t)\rangle}_i)+\xi_i
\label{eq2}
\end{equation}

${\langle u(t)\rangle}_i$ is the local average of directions of particles and disorders around the 
 {\em$i^{th}$} particle\cite{czirok1999collective}. It is calculated by averaging over  the particles and disorder in the interval $[x_i-\Delta,x_i+\Delta]$ with $\Delta=1$. $\xi_i\in[-\frac{\eta}{2},\frac{\eta}{2}]$. $v_0$ is the magnitude of the self-propulsion speed and it is fixed to $1$
\begin{equation}
    G(u) = \begin{cases} 
      \frac{u+1}{2} & for \qquad {u  \geq 0} \\
      \frac{u-1}{2} & for \qquad  {u  < 0} \\
    \end{cases}
   \label{step}
\end{equation}
In the presence of disorder, the 

\begin{equation}
    {\langle u\rangle}_i=\sum{}{} ((u_j+\beta {u_j}')/n_r)
    \label{eq4}
\end{equation}

\begin{figure} [hbt]
{\includegraphics[width=1.0 \linewidth]{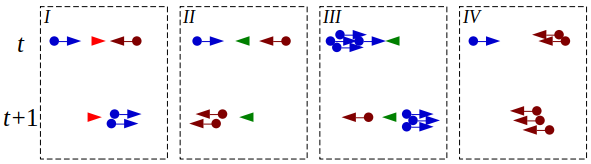}}
	\caption{In this illustrative diagram, all the possible scenarios of the flipping of SPPs are showcased in different panels. (I-III) represent when there is a disorder in the path of SPPs, (IV) when there is no disorder is present. SPPs and disorders are depicted as a circular entity connected with an arrow, left-moving SPPs are marked in blue, while right-moving ones are portrayed in brown. Similarly, quenched disorders within the system are illustrated with an arrowhead; left and right-oriented disorders are marked in green and red, respectively. The SPPs and disorders in the top row are shown at a particular time $t$ before the interaction with SPPs/disorders and in the bottom row at time $t+1$ after the interaction with SPPs/disorders.  }
\label{fig:1}
\end{figure}

$u_j$ and ${u_j}'$ are orientations of particles and disorders respectively in the interaction radius of $i^{th}$ particle. $\beta = 10$ the interaction strength of particles with disorder. $n_r$ is the number of particles and disorders in the interval $[x_i-\Delta, x_i+\Delta]$.
The disorders which have either left or right orientation, try to reorient the particles in their direction. The cartoon of the effect of orientation interaction;  the first term on the right hand side of Eq. \ref{eq2} for four different scenarios is shown in Fig.\ref{fig:1}. Fig.\ref{fig:1}(I)-(II) shows the disorder acting like transmitter/reflector when the direction of incoming particle  is same/opposite to the direction of disorder. Fig.\ref{fig:1}(III)shows the scenario when a cluster of particles interact with another particle and no disorder is present at that location, then the majority wins and a bigger cluster is formed. Fig.\ref{fig:1}(IV) shows a big cluster of particles interacting with disorder with the  direction of disorder  opposite to the direction of incoming cluster,causing the big cluster to split and a fraction of particles from the cluster getting reflected and another part getting transmitted.  For clean system or for zero disorder $\rho_d = 0$, model reduces to the  model introduced by Czirok A {\em et al.} \cite{czirok1999collective}. We name this case as {\em pure system}.

{\em Coarse Grained Model:-} We further introduce the coarse-grained  model for the collection of self propelled particles with quenched disorders in one dimension. We define the coarse-grained fields by dividing  the whole system in  units of  size $2 \Delta$. The local coarse-grained density and magnetisation is defined as  $\rho(x, t) = \frac{N_x(t)}{2 \Delta}$, where $N_x(t)$  is the number of particles at the lattice point $x$ in the interval $[x-\Delta, x+\Delta]$  and $m(x, t)$ is obtained by taking average orientation of all the particles in the same interval at time $t$. The effect of disorder is added as an additional noise which is random in space and quench in time.  The equations of motion for the coarse-grained density $\rho(x, t)$ and magnetisation are given by $m(x, t)$


\begin{equation}
\partial_t\rho = D\partial_{x}^2\rho-v_0\partial_x(m\rho)
\label{eq5}
\end{equation}
\begin{equation}
\partial_tm=D\partial_{x}^2m-v_0\partial_x(\rho)+(\alpha(\rho)-\alpha_2m^2)m+f_m+\xi \\
\label{eq6}
\end{equation}
The first terms on the right hand side of Eq.\ref{eq5} is the diffusion in density, the second term is the term due to self-propelled nature of particle, where the active current is proportional to the local magnetisation $m$. The first term on the right hand side of Eq. \ref{eq6} is again the diffusion in magnetisation, the second term is the coupling due to density inhomogeneity and present due to self-propelled nature of particles. The third term is usual order-disorder term, where region with high density promotes ordering and regions with low density favours random orientation. Such density dependent order-disorder coupling is introduced by taking $\alpha(\rho) = (\frac{\rho}{\rho_c}-1)$. Hence it changes sign at $\rho = \rho_c$. $\alpha_2$ is a positive constant. The second last term $f_{m}$ is the random annealed Gaussian white-noise with mean zero and variance  $\Delta_m$, such that $\langle f_m(x, t) f_m(x', t')\rangle = 2 \Delta_m \delta(t-t')\delta(x-x')$. This term is present due to randomness in the system, which acts like thermal noise in the coarse-grained model. The last term is the effect of disorder and it is introduced as an additional Gaussian white-noise; quenched in time and  random in space with mean zero and variance $\Delta_d$, such that 
$<\zeta(x) \zeta(x')> = 2\Delta_d \delta(x-x')$. Hence $\Delta_d$  serves as the strength of noise, which is varied from ($0, 0.004-0.007$). For $\Delta_d=0$, the model reduces to the clean system and it is analogous to the one-dimensional limit of the {\em Toner-Tu} equation for polar flock \cite{tonertupre1998,Toner1995}. 
\begin{figure*}[hbt]
    \centering
    \includegraphics[width=0.98\textwidth]{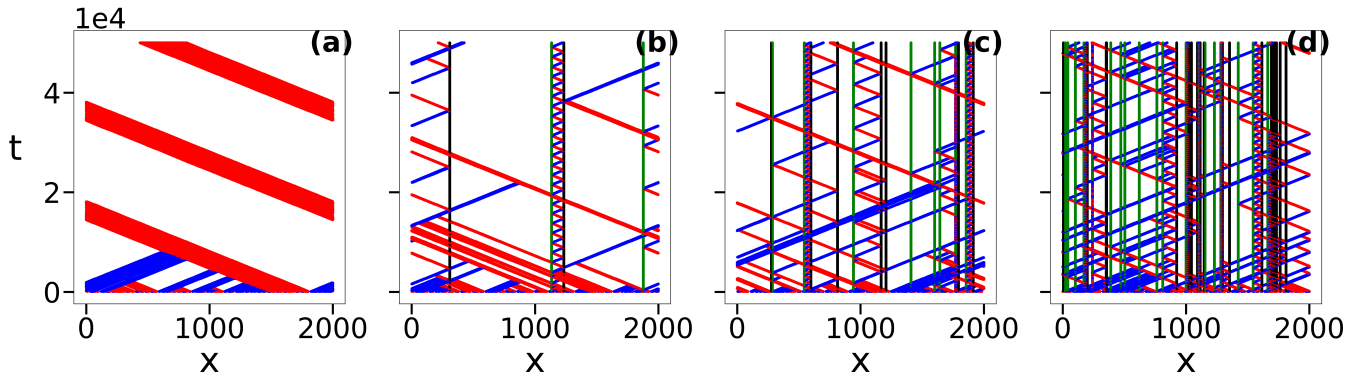}
    \caption{In this series of spatio temporal snapshots, the one-dimensional system unfolds over time, where the x-axis represents the spatial coordinates of SPPs and disorders. The y-axis signifies the progression of time. SPPs with leftward orientation are depicted in red, while those with rightward orientation are shown in blue. Left-oriented obstacles are represented in green, and right-oriented obstacles in brown. These visualizations capture the dynamic interplay between SPPs and immobile obstacles, illustrating how varying obstacle densities influence the emergence of ordering and clustering phenomena within the system.}
    \label{fig3}
\end{figure*}
\begin{figure*}[hbt]
    \centering
    \includegraphics[width=0.98\textwidth]{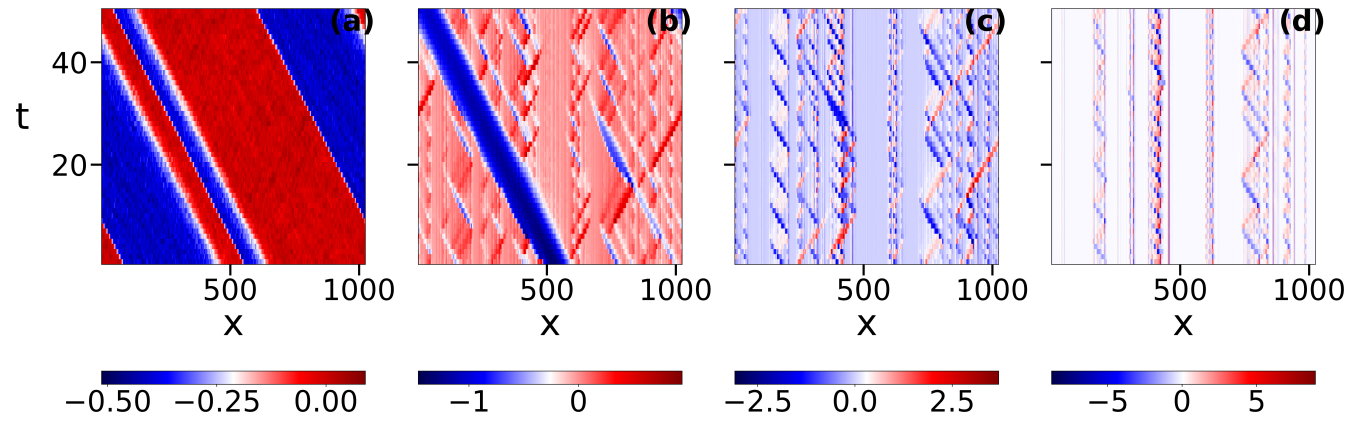}
    \caption{The spatio-temporal snapshots (a-d)  illustrates the effect of increasing the disorder strength $\Delta_d = (0, 0.001, 0.01, 0.04)$ respectively in the coarse-grained model. The $x-$axis denotes the product of local magnetization and density  $g(x,t)$ and the $y-$axis represents the increasing time direction.  The color bar represents the value of $g(x,t)$}
    \label{fig2}
\end{figure*}
\section{Results}
\label{result}
To understand the effect of disorder in the system we first plot the spatio-temporal plot of local density and magnetisation for different densities $\rho_d$ and strengths $\Delta_d$ of disorder in microscopic and coarse-grained models respectively. Fig. \ref{fig3}(a-d) shows the spatio-temporal plot of particle's position in the microscopic model. The two colors show the two types of orientation of particles; red color represents the left orientation and blue color shows the right orientation. The first plot in Fig. \ref{fig3}(a) is for the zero disorder or clean system and shows  the macroscopic ordering in the system. The vertical direction shows the flow in time and horizontal axis is the system dimension. It is very clear that for the clean system initially there are clusters of different orientations and with time the clusters merge and a big macroscopic cluster is formed  and moves coherently in the direction of its orientation. The direction of motion of the cluster is spontaneously chosen in the system. \\
The Fig. \ref{fig3}(b-d) shows the same plot in the presence of disorder and for different disorder densities. Fig. \ref{fig3}(b) is for small density of disorder $\rho_d = 0.001$. It is clear that the disorder affects both the ordering and clustering in the system. The macroscopic clustering in the system is broken and big cluster splits in clusters of different orientation and  high density (narrow bands). The direction of clusters very much depends on the local orientation of disorder. On further increasing disorder as shown in Fig. \ref{fig3}(c-d), the cluster of particles splits in many small clusters and many times  these clusters perform the back and forth motion between two oppositely oriented disorder.\\

Similarly we make the spatio-temporal plot of the local magnetisation density $g(x, t) = \rho(x, t)m(x, t)$ for the coarse-grained model for different strengths of disorder $\Delta_d$. The corresponding snapshots are shown in Fig.\ref{fig2}(a-d) for four different disorder strengths including clean system $\Delta_d = 0$. Clearly for the clean system as shown in Fig.\ref{fig2}(a), macroscopic clustering is observed with one big cluster moving in the direction of local magnetisation. Color bars in Fig.\ref{fig2} show the value of $g(x, t)$ which denotes the variation of the local magnetisation density. If $g(x)>0$, majority particles are moving in right direction while for $g(x)<0$, they are moving towards left.  As we increase the strength of disorder macroscopic clustering breaks and small denser clusters are formed, as can be seen by the range of the color bar for different strengths. Increasing disorder leads to increase in particle density in the clusters rendering them immobile. 
Hence disorder not only splits the clustering and breaks the ordering but also creates the localisation of particles. \\

\begin{figure} [hbt]
\includegraphics[width=0.98 \linewidth]{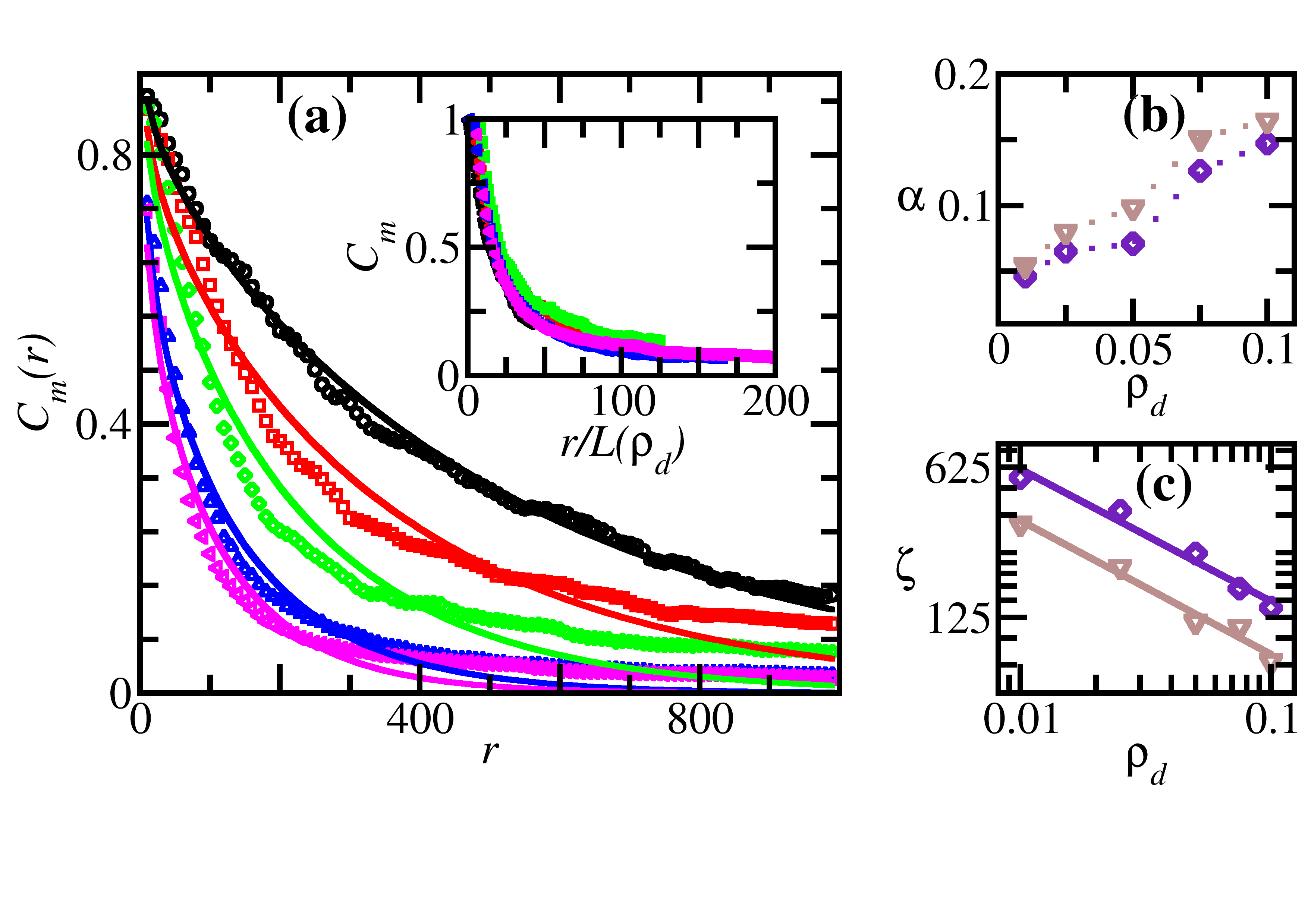}
\caption{(a) Two-point Spatial correlation function $C_m(r)$ {\em vs.} $r$ for different disorder densities ($\rho_d$ = 0.010, 0.025, 0.050, 0.075 and 0.100) for microscopic model in the steady state at time $t=12000$. The symbols are data from numerical simulation and lines are fit to the proposed form of the $C_m(r)$ as given in Eq. \ref{eq1}. {\em Inset}: The plot of scaled $C_m(r$ {\em vs.} scaled distance $r/L(\rho_d)$ for different disorder densities. Plot (b) shows the power exponent $\alpha(\rho_d)$ {\em  vs.} $\rho_d$ on linear scale. The two symbols show the power exponent for different packing fraction of the SPPs/disorders along the fixed length, violet square is for packing 1.0 while brown triangle for packing 2.0. (c)  Demonstrates the cluster size $\zeta(\rho_d)$ {\em vs.}  $\rho_d$ on $\log-\log$ scale. Symbols have usual meaning as in (b).}
\label{fig5}
\end{figure}

\begin{figure} [hbt]

\includegraphics[width=0.98 \linewidth]{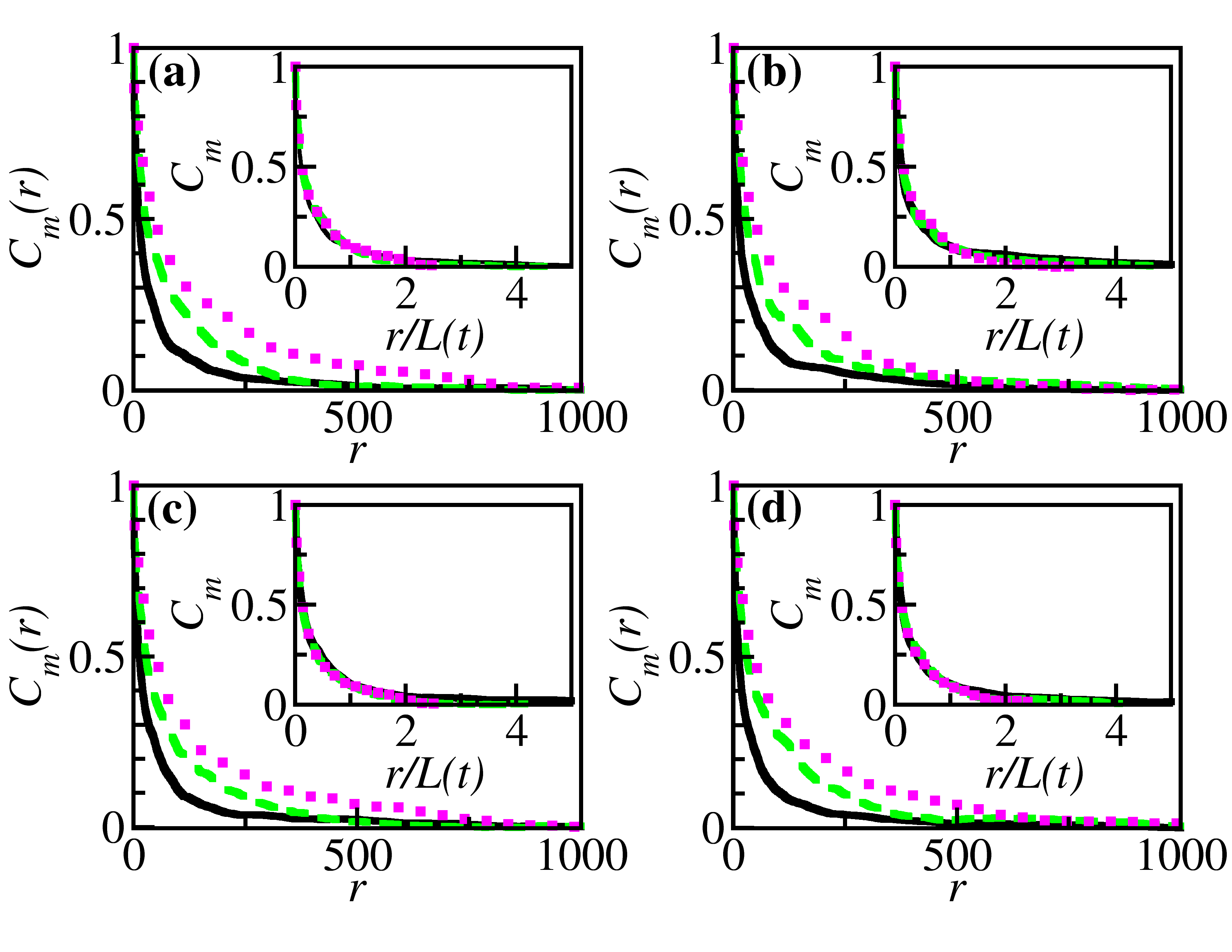}

	\caption{Two point correlation function $C_m(r)$ {\em vs.} $r$ at different times $t=500, 1000, 1500$ shown by black (solid line) green (dashed line) and mazenta (dotted line) respectively and  disorder densities $\rho_d = 0.010, 0.025, 0.050$ shown in different panels from $a-d$ respectively. {\em Insets}: The scaled $C_m(r)$ {\em vs.} scaled distance $r/L(\rho_d)$ for all the fourcases.}
\label{fig6}
\end{figure}

\begin{figure} [hbt]

\includegraphics[width=0.99 \linewidth]{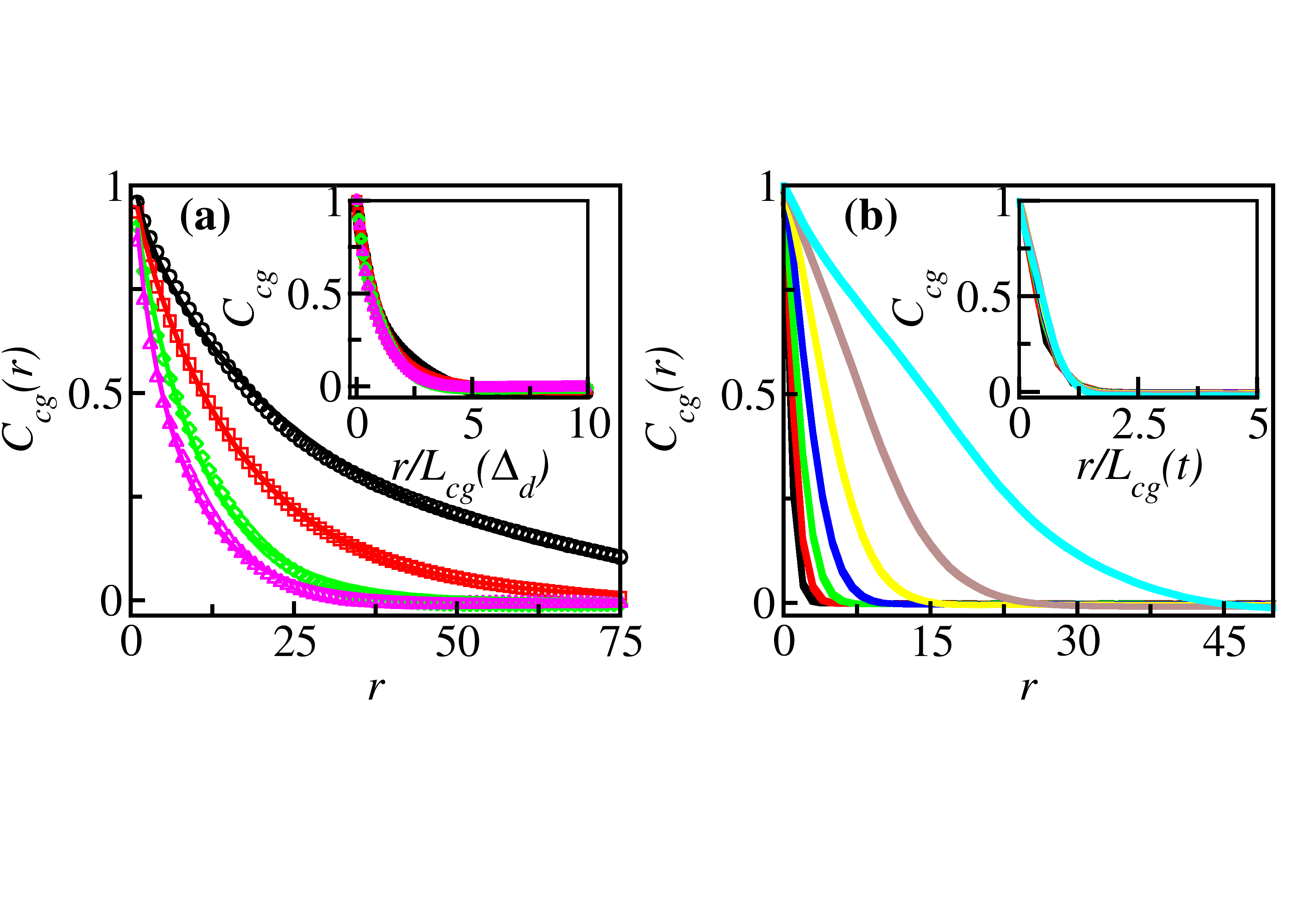}
\caption{(a) Shows the plot of two-point correlation function $C_{cg}(r)$  for various disorder strengths, $\Delta_d$= $0.001$ (black, circles), $0.002$ (red, squares), $0.004$ (green, diamonds), $0.007$ (pink, triangles). {\em Inset} : The scaled $C_cg(r)$ {\em vs.} scaled distance $r/L(t)$. (b) shows the correlation function at different times for a particular disorder strength $\Delta_d = 0.001$. {\em Inset}: The scaled $C_{cg}(r)$ {\em vs.} scaled distance $r/L(t)$.}
\label{fig4}
\end{figure}

Till now we have discussed the impact of disorder in terms of spatio-temporal snapshots of the system. We further characterise the system by calculating two-point magnetisation correlation function.  We will first focus on the results for two-point correlation function obtained from the microscopic model. The two-point magnetisation correlation function  is defined as $C_m(r) = \langle u_iu_j\rangle_{r}$ where the $i$ and $j$ are the indices for pair of particles  and $\langle .. \rangle_r$, means sum over all the ($i$, $j$) pairs  with the separation $r$ and $r\pm \Delta$. Further the correlation is averaged over 40 independent realisations. In Fig. \ref{fig5} (a) we show the variation of $C_m(r)$ vs. distance $r$ for different disorder densities { $\rho_d = 0.010, 0.025, 0.050, 0.075$ and $0.100$} and particles density $\rho = 1.0$ in the steady state. As shown in the snapshot in Fig.\ref{fig3} (a) and (b), the disorder breaks the one big macroscopic cluster in smaller clusters, hence long-ranged ordering is destroyed. In Fig.\ref{fig5}(a), we show the $C_m(r)$ vs $r$ for different disorders and it decays with distance $r$. Increasing disorder lead to sharp decay of $C_m(r)$ as a function of distance. 
We  analysed the form of two-point correlation function and found that it is neither a pure exponential nor a pure power-law. Although we find it a combination of algebraic  and exponential decay. Hence we propose a form for the two-point correlation function 
\begin{equation}
C_m(r) = \frac{1}{r^{\alpha(\rho_d)}}\exp(-r/\zeta(\rho_d))
\label{eq1}
    \end{equation}
    where the exponent $\alpha(\rho_d)$ for the power-law and the cluster size $\zeta(\rho_d)$ are functions of disorder density $\rho_d$. We obtained the $\alpha(\rho_d)$ and $\zeta(\rho_d)$ by fitting the $C_m(r)$ with the proposed form. In Fig.\ref{fig5}, lines are fitting functions and data points are obtained from numerical simulation. For all the cases we find the good fit of the $C_m(r)$ with respect to the proposed form of the correlation. In the  Fig. \ref{fig5} (b-c) we show the variation of $\alpha(\rho_d)$ and $\zeta(\rho_d)$ {\em vs.} $\rho_d$ on linear and $\log-\log$ scale for two densities $\rho = 0.5$ and $2.0$. It is very clear that the $\alpha(\rho_d)$ increases linearly with disorder density and $\zeta(\rho_d)$ decays algebraically with $\rho_d$ with $\zeta(\rho_d) \simeq \frac{1}{{\rho_d}^{\beta}}$. From the fit we found $\beta \simeq 0.61$.   We further check the results for other particle densities. For all the densities we find the good fit of the $C_m(r)$ with respect to the proposed form for the correlation (data not shown). Further  we also check the static scaling of the two-point correlation function by showing the scaled plot of $C_m(r)$ vs. scaled distance $r/L({\rho_d})$ in Fig. \ref{fig5}(a)({\em inset}), where $L(\rho_d)$ is obtained by $0.1$ crossing of $C_m(r)$. Very clearly good static scaling is observed for all disorder densities. Such scaling of data suggest the scale invariance of two-point correlation function for all disorder densities. \\
    Now we check the ordering kinetics and dynamic scaling of the two-point correlation function using microscopic simulation. In Fig. \ref{fig6}(a-d)
we show the plot of $C_m(r)$ {\em vs.} $r$ for four different disorder densities $\rho_d=0.010, 0.025, 0.050, 0.100 $ respectively and three different times $t=500, 1000, 1500$. For all disorder densities, the correlation increases with time, which suggest the growth of clusters. Further the insets of Fig. \ref{fig6}(a-d) shows the scaling collapse of the same data as shown in the main panel. The correlation length is obtained by $0.1$ crossing of the $C_m(r)$. For all disorder densities  we find good dynamic scaling.    \\
Further we also compared the two-point correlation function calculated in the coarse-grained simulation $\mathcal{C}_{cg}(r) = \langle\delta m(r_0, t) \delta m(r+r_0, t)\rangle_{r_0}$, where $\langle .. \rangle$ means average over many reference points and over $100$ independent realisations.  $\delta m(r, t)$ is the fluctuations in $m(r, t)$ from the mean value. Where mean is calculated by averaging over all points. In Fig. \ref{fig4}(a-b) we show the plot of $\mathcal{C}_{cg}(r)$ vs. distance $r$ for different strengths of disorders $\Delta_d = 0.001, 0.002, 0.004$ and $0.007$ in the steady state and for disorder strength $0.001$ at different times respectively. In Fig. \ref{fig4}(a), the lines are the fit of the correlation function with the combination of power-law and exponential as proposed in Eq. \ref{eq1} and symbols are from the coarse-grained numerical simulations. For a fixed time and increasing disorder the correlation steepens Fig. \ref{fig4}(a) and for fixed disorder and increasing time correlation increases with time Fig. \ref{fig4}(b). Hence the size of the ordered region increases with time. We calculated the correlation length $L(t)$ by the first $0.1$ crossing of $\mathcal{C}_{cg}(r)$ and in the insets of Fig. \ref{fig4}(a-b) we show the scaled correlation function $\mathcal{C}_{cg}(r)$ {\em vs.} scaled distance $r/L(t)$. We find  the good static and dynamic scaling in the system for disorder strengths $\Delta_d$. \\

\section{Conclusions}

We study the effect of quenched disorder on the characteristic of self-propelled particles (SPPs)
in one-dimension. The SPPs, characterized by distinct left or right orientations interact with disorder which serve as directional cues. Through extensive microscopic simulations and a numerical exploration of coarse-grained equations of motion for local density and orientation of particles,  our investigation reveals how the density of the disorder influence the emergence of ordering and clustering in the collection of the SPPs. Quantifying the degree of ordering, we measure the characteristic correlation length and its corresponding exponent across varying disorder densities. The disorder affects the macroscopic ordering in the system, the size of the ordered clusters decays algebraically with
disorder. Further the disorder also affects the clustering of particles, in the presence of disorder a big macroscopic cluster breaks in small clusters resulting in the increase in the number of clusters as obstacle density escalates and lead to the localisation of particles around it and ultimately resulting in high density around the disorder. We observe  identical results from the microscopic and coarse grained model. \\
This model can be extended to quasi one-dimension system to study the effect of disorder on the ordering of SPPs at the interface of crossover in the dimensions. Our  study can help in the  understanding  of the confinement of mesenchymal cells  to thin adhesive one-dimensional (1D) lines with a characteristic spatial force
pattern \cite{hennig2020stick}.

\label{conclusion}

\bibliography{citation}

\end{document}